\documentclass[aps,pra,twocolumn,preprintnumbers]{revtex4}
\usepackage{graphicx}
\usepackage{amssymb}
\usepackage{amsmath}
\newcommand{\be}{\begin{equation}}
\newcommand{\ee}{\end{equation}}
\newcommand{\bea}{\begin{eqnarray}}
\newcommand{\eea}{\end{eqnarray}}
\newcommand{\Eq}[1]{Eq.\,(\ref{#1})}
\newcommand{\Fig}[1]{Fig.\,\ref{#1}}

\begin{document}
\title{Comment on ``Normalization of quasinormal modes in leaky optical cavities and plasmonic resonators''}
\author{E.\,A. Muljarov}
\author{W. Langbein}
\affiliation{School of Physics and Astronomy, Cardiff University, Cardiff CF24 3AA,
United Kingdom}
\begin{abstract}
Recently, Kristensen, Ge and Hughes have compared [Phys. Rev. A {\bf 92}, 053810 (2015)] three different methods for normalization of quasinormal modes in open optical systems, and concluded that they all provide the same result. We show here that this conclusion is incorrect and illustrate that the normalization of [Opt. Lett. {\bf 37},  1649  (2012)] is divergent for any optical mode having a finite quality factor, and that the Silver-M\"uller radiation condition is not fulfilled for quasinormal modes.
\end{abstract}
%
%
\date{\today}
\maketitle

In a recent paper~\cite{KristensenPRA15}, Kristensen {\em et al.} have considered three different normalizations of quasinormal modes: (i) the normalization given in~\cite{KristensenOL12}, which is a generalized version of the work by Leung {\em et al.}~\cite{LeungJOSAB96} and thus called here {\em Leung-Kristensen (LK)}; (ii) the normalization introduced in~\cite{MuljarovEPL10},  which is analytically {\em exact}; and (iii) the normalization suggested in \cite{SauvanPRL13}, based on perfectly matched layers ({\em PML}). Kristensen {\em et al.} concluded that all three normalizations provide the same result. We show in this Comment that (i) the LK normalization is divergent, and therefore ill-defined. A regularized variant of the LK normalization, put forward in~\cite{KristensenPRA15}, is not suited for numerically determined resonant states (RSs); (ii) the claimed equivalence of LK and PML normalizations is incorrect since the Silver-M\"uller radiation condition used in the argumentation is not valid for RSs.

The LK normalization, Eq.\,(5) of~\cite{KristensenPRA15}, for an optical system surrounded by vacuum is defined by an infinite-volume limit
\be
{N}^\infty_{\rm LK}=\lim_{V \to \infty} N_{\rm LK}
\label{LK1}
\ee
of the normalization
\be
N_{\rm LK}=\int_V \varepsilon ({\bf r}) {\bf E}^2 ({\bf r}) dV + \frac{i}{2k}\oint_{S_V} {\bf E}^2 ({\bf r}) dS\,,
\label{LK2}
\ee
calculated over the finite volume $V$ and its surface $S_V$, using the electric field ${\bf E}({\bf r})$ and the wave vector $k$ of the quasinormal mode, which we call here {\em resonant state}, adopting its original name in the literature \cite{Garcia76}.  Let us assume for now that the volume is a sphere of radius $R$ with the system in its center.

We first show that $N_{\rm LK}$ diverges for $R \to \infty$, so that the LK normalization ${N}^\infty_{\rm LK}$ mathematically does not exist. The dependence of $N_{\rm LK}$ on $R$ was evaluated in~\cite{KristensenPRA15} by expanding ${\bf E}({\bf r})$ into vector spherical harmonics, with the spherical harmonics $Y_{lm}(\theta,\varphi)$ and Hankel functions of first kind $h_l(z)$ as basis (here $l$ is the orbital quantum number). Since $k$ is complex for any RS having a finite quality factor ($Q$-factor) $Q=|{\rm Re}(k)/[2{\rm Im}(k)]|$, the argument of $h_l(z)$ is also complex: $z=kR$. The limiting form of $i^{l+1} h_l(z)\to {e^{iz}}/{z}$ given in Eq.\,(9) of~\cite{KristensenPRA15} neglects diverging contributions, since the exact form is given by
\be
i^{l+1} h_l(z)=\frac{e^{iz}}{z}P_l(\xi)\,,
\label{Hankel}
\ee
where
\be
P_l(\xi)=\sum_{m=0}^l \frac{(l+m)!}{(l-m)!m!} \,\xi^m\ \ \ {\rm and}\ \ \ \xi=\frac{1}{-2iz}\,.
\ee
Now, $P_l(\xi)$ is a polynomial of order $l$, and all resulting terms of \Eq{Hankel} diverge for complex $z$, owing to the exponentially large factor $e^{iz}$. Consequently, Eq.\,(10) in~\cite{KristensenPRA15}, based on Eq.\,(9) and stating that $\partial_R \hat{I}^r_l(R)=0$, is incorrect, and should read instead
\bea
\partial_R \hat{I}^r_l(R)&=&R^2 h_l(z)\left[h_l(z)+i h'_l(z)+i \frac{h_l(z)}{z}\right]\nonumber
\\
&=& \frac{ h_l^2(kR)}{2k^2}\,\frac{P_l'(\xi)}{P_l(\xi)}.
\label{I_l}
\eea
In particular, $ P_l'(0)/P_l(0)=l(l+1)$, and thus $\partial_R \hat{I}^r_l(R)=0$ holds only for $l=0$. However, electromagnetic modes with $l=0$ do not exist in finite three-dimensional optical systems. Therefore, in general, $\partial_R \hat{I}^r_l(R)\to\infty$ for $R\to\infty$. For example, considering $l=1$ we find
\be
\partial_R \hat{I}^r_1(R)= \frac{e^{2ikR}}{k^4 R^2}\left(1+\frac{i}{kR}\right).
\ee
The authors of \cite{KristensenPRA15} write ``{\em In practice, direct application of Eq.\,(5) leads to an integral
that seems to quickly converge towards a finite value, but in fact
oscillates about this value with an amplitude that eventually
starts to grow (exponentially) with the distance, albeit slowly
compared to the length scales in typical calculations. This was
noted in Ref. [5], where the oscillations were observed only for
the cavity with the lowest quality factor ($Q\approx16$).}'' In the cited reference \cite{KristensenOL12}, we find ``{\em For very low-Q cavities, however, the convergence is nontrivial
due to the exponential divergence of the modes that
may cause the inner product to oscillate around the proper
value as a function of calculation domain size}'', and otherwise  ``{\em quick convergence}'' is claimed. The residual $f^{\rm res}_{\rm LK}(R)$ of the LK normalization, which is given in Eq.\,(11) of~\cite{KristensenPRA15} diverges -- its precise form is
\bea
f^{\rm res}_{\rm LK}(R)&=&\frac{R^3}{2}\left[h_l^2(z)-h_{l-1}(z)h_{l+1}(z) +\frac{i}{z} h_l^2(z)\right]
\nonumber
\\
&=&\frac{e^{2ikR}}{k^5 R^2}\, Q_{2l-2}\left(\xi\right)\,,
\label{residual}
\eea
where $Q_n(\xi)$ is an $n$-th order polynomial of $\xi=(-2ikR)^{-1}$, with the leading term at small $\xi$ (i.e. at large $R$) given by
$Q_{2l-2}(0)=-i(-1)^{l+1} {l(l+1)}/{2}$\,, see~\cite{MuljarovARX14} for more details. Therefore, $N_{\rm LK}\to\infty$ as $R\to\infty$.

The authors of ~\cite{KristensenPRA15} describe this divergence as ``{\em Thus, while Eqs.(9) and (10) appear to be formally correct also for complex arguments, the limit $R\to\infty$ in practice leads to a position dependent phase difference between the Hankel function and its limiting form, which makes the limit nontrivial to perform along the real axis.}''. We note that (i) there is no difference between formalism and practise in mathematical limits; (ii) the limit $V\to\infty$ is defined for real volumes, and thus real $R$; (iii) the limit of $N_{\rm LK}$ along the real axis of $R$ is not ``nontrivial'', it simply does not exist due to the divergence.

\begin{figure}
	\includegraphics*[width=0.98\columnwidth]{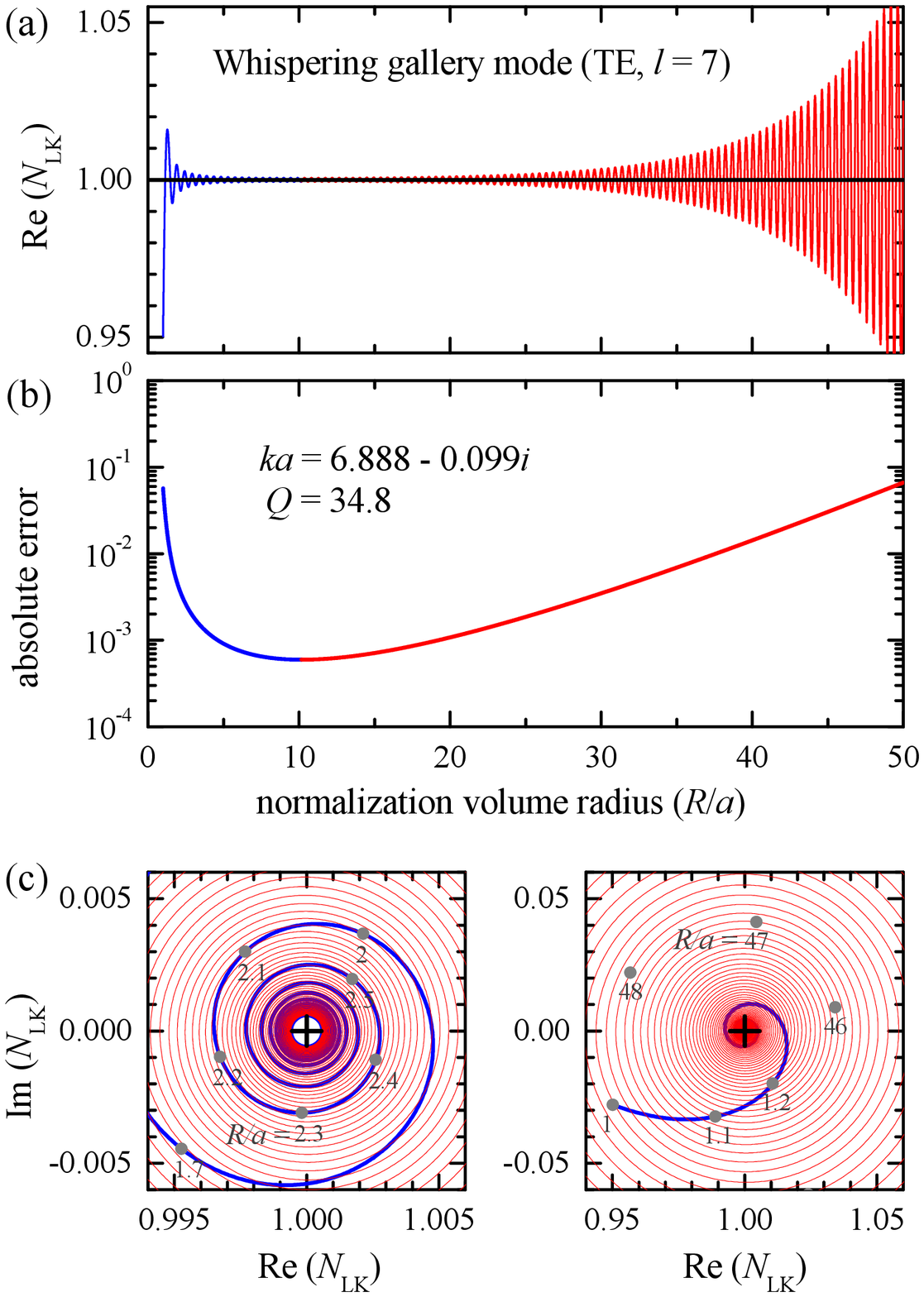}
	\caption{ LK normalization $N_{\rm LK}$ (a,c-d) and its absolute error $|N_{\rm LK}-1|$ (b) as a function of the radius $R$ of the spherical volume, for a TE $l=7$ WGM of a dielectric sphere of refractive index $n_r=2$ and radius $a$, in vacuum. The wave vector of the WGM is $ka=6.888-0.099\,i$, corresponding to $Q=34.8$. Blue (red) color shows the region of error decreasing (increasing) with $R$. The exact normalization is shown by a black line (a) and a black cross (c-d).
	}
	\label{fig:WGM}
\end{figure}

\begin{figure}
\includegraphics*[width=0.98\columnwidth]{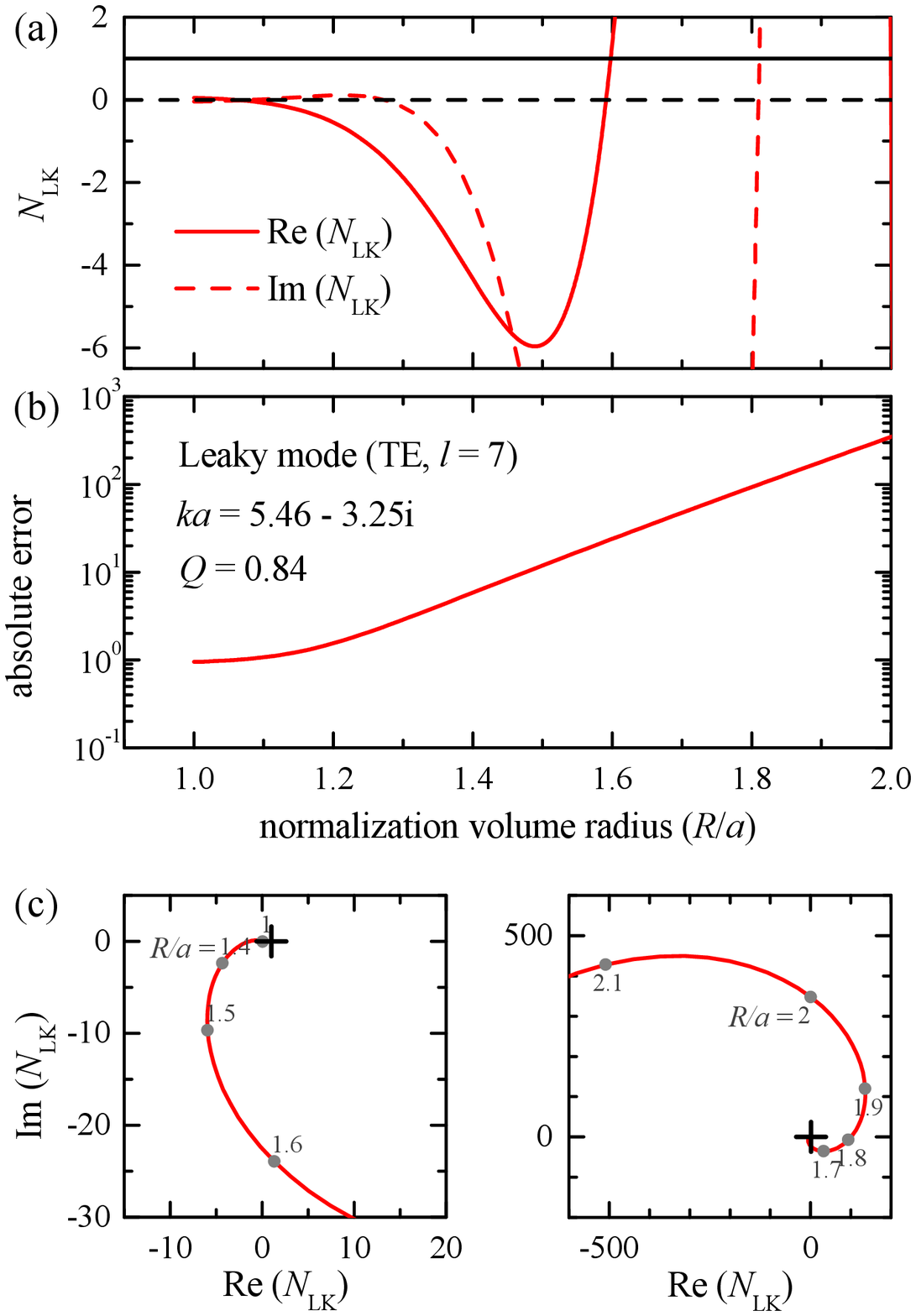}
\caption{ As in Fig.\,\ref{fig:WGM} but for a leaky mode with the wave vector $ka=5.46-3.25\,i$, corresponding to $Q=0.84$.
}
\label{fig:LM}
\end{figure}

\begin{figure}
\includegraphics*[width=0.98\columnwidth]{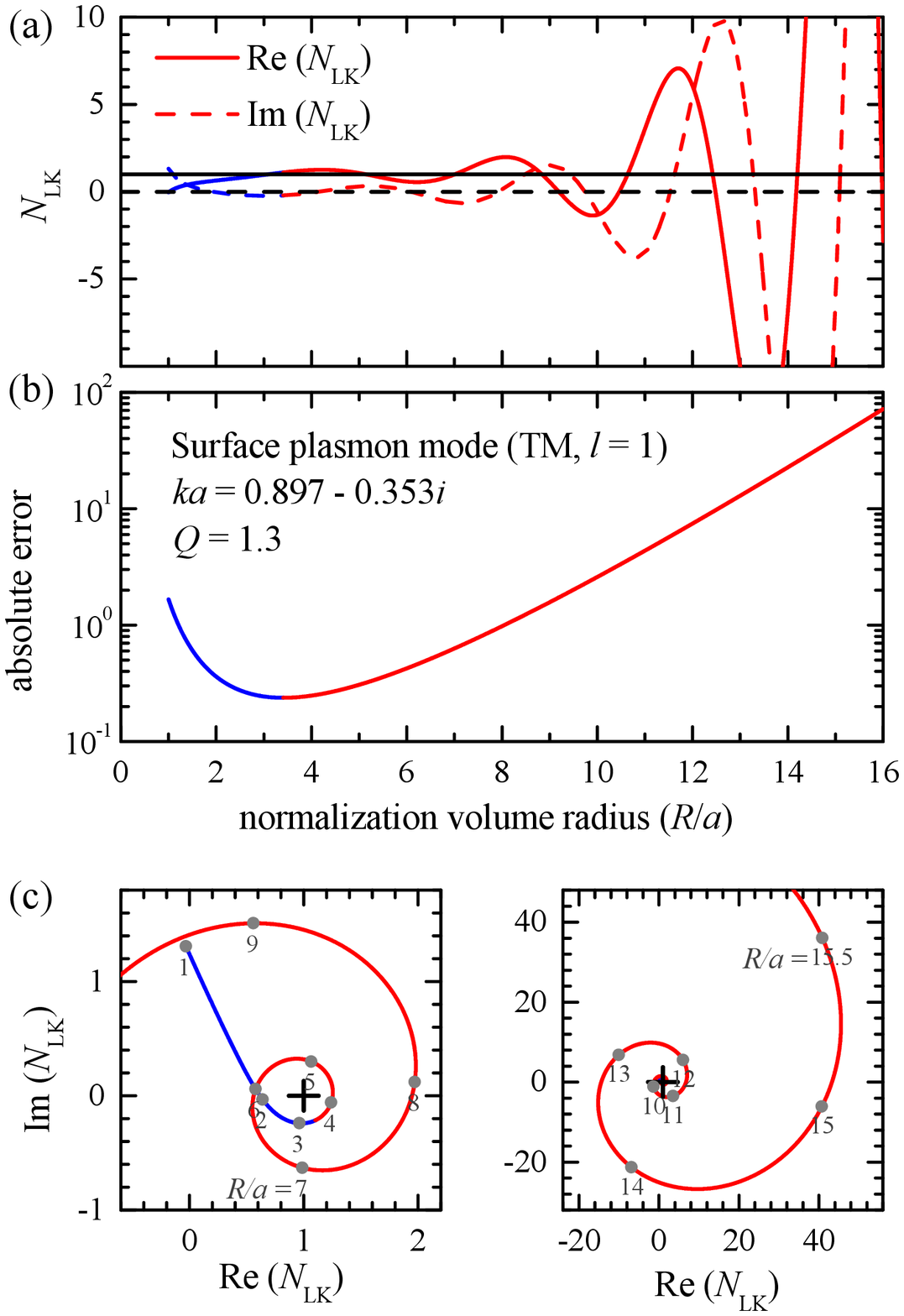}
\caption{ As in Fig.\,\ref{fig:WGM} but for a surface plasmon $l=1$  transverse magnetic (TM) mode in a gold sphere in vacuum. The mode wave vector is $ka=0.897-0.353\,i$,  corresponding to $Q=1.27$. The radius of the sphere is $a=100$\,nm, and the gold permittivity was treated using a Drude model~\cite{SauvanPRL13}.
}
\label{fig:SP}
\end{figure}

We show in Figures \ref{fig:WGM}-\ref{fig:SP} the $R$-dependence of $N_{\rm LK}$ for RSs of a dielectric sphere of radius $a$ with high and low $Q$-factors, and for the fundamental plasmonic RS of a gold sphere. All RS fields used have been normalized using the exact normalization, having analytical expressions \cite{DoostPRA14,MuljarovARX14}. We commence using a RS with a $Q$-factor of about 35 (similar to the RS illustrated in Fig.\,3 of~\cite{KristensenPRA15}), the $l=7$ transverse electric (TE) whispering gallery mode (WGM) of a dielectric sphere with refractive index $n_r=2$ in vacuum. \Fig{fig:WGM} is formatted similarly to Fig.\,3 of~\cite{KristensenPRA15}, showing in blue the $R$-region of convergence (spiralling in), and in red the $R$-region of divergence (spiralling out) of $N_{\rm LK}$ in the complex plane. We note that the spiralling out region is not shown in Fig.\,3 of \cite{KristensenPRA15}.

One could argue that for high-$Q$ modes, the LK normalization can be sufficiently accurate, as the error reaches $10^{-3}$ at $R\approx10a$ in the present example. One could even refine this result by evaluating the center of the spiral, as suggested in~\cite{KristensenPRA15}. However, one has to keep in mind that simulating the required extended spatial domain in numerical calculations is computationally costly. On the other hand, evaluating the LK normalization close to the system, leads to significant errors due to the slow $1/R^2$ dependence of the residual term \Eq{residual}, as is clearly shown by the blue line in Figs.\,\ref{fig:WGM}(b) and (d). The LK normalization used for high-$Q$ RSs is thus at least inconvenient, due to the large computational domain required to obtain sufficient accuracy. More discussion and data are given in the supplement of~\cite{MuljarovARX14}.

A RS with a low $Q$-factor of about 1 in the same dielectric sphere, a TE $l=7$ leaky mode, is used in Fig.\,\ref{fig:LM}. We see that $N_{\rm LK}$ starts close to zero at $R=a$ and then spirals out in the complex plane. This results in an initial error of about 100\%, increasing to 40000\% at $R=2a$, prohibiting to extract a value for the LK normalization.

Finally, we show in \Fig{fig:SP} the LK normalization of the fundamental surface plasmon mode of a nano-plasmonic resonator -- a gold sphere in vacuum, 200\,nm in diameter, also used in \cite{SauvanPRL13}. This mode has a $Q$-factor of about 1.3. There is an initial decrease of the error from 200\% down to about 10\%, followed by an exponential divergence. A single loop in the complex plane is observed, circling the correct normalization. The minimum of the error is observed at about $R=3a$, thus requiring a much larger computational domain than the system size. A reliable extraction of the RS normalization from $N_{\rm LK}$ in this case is questionable.

A regularized version of the LK normalization suggested in~\cite{KristensenPRA15} is
based on an analytic continuation of the electric field into the complex plane of $R$ and taking a limit of $R\to-\infty$.
For this to be used, the fields of RSs have to be known analytically. This regularization is thus not suited for numerically determined RSs.
We emphasize that this ``regularized'' LK normalization is a different quantity compared to the divergent LK normalization defined by Eqs.\,(\ref{LK1}-\ref{LK2}) what  was actually used in~\cite{KristensenOL12} and in numerous follow-up publications of the same group, including the numerical examples of~\cite{KristensenPRA15}.

The exact normalization~\cite{MuljarovEPL10} is independent of $V$ and differs from the LK normalization only by the surface term. To understand the physical difference between the surface terms, we consider a small piece $\Delta S$ of the surface of integration and assume for simplicity that the local electric field of the RS has the form of a plane wave ${\bf E}={\bf E}_0 e^{i{\bf k}\cdot {\bf r}}$ propagating in the direction of ${\bf k}$, with ${\bf k}^2=k^2$ and a constant amplitude ${\bf E}_0$. Then, after simple algebra, we find that the selected part of the surface integral in the exact normalization is given by
\be
\frac{i}{2k^2}\int_{\Delta S} E^2\, ({\bf k}\cdot\hat{\bf n}) dS\,,
\label{surf}
\ee
where $\hat{\bf n}$ is the surface normal, while for the LK normalization the corresponding part is
\be
\frac{i}{2k}\int_{\Delta S} E^2\, dS\,.
\label{surf_LK}
\ee
This shows that the LK surface term assumes that the propagation direction of the field is always normal to the surface, while the exact normalization takes the actual propagation direction into account. The two terms are equal only if $\hat {\bf n}\parallel {\bf k}$ over the whole surface, which is not possible in electrodynamics due to the vectorial nature of the electro-magnetic field.

The implicit assumption of normal outward propagation makes the LK normalization not only diverging for $V\to\infty$, but also depending on the surface shape. Note that the shape of $S_V$ in the LK normalization is not restricted to spherical surfaces, and a cuboid was actually used in one of the examples shown in~\cite{KristensenOL12} and~\cite{KristensenPRA15}. However, since the surface term in $N_{\rm LK}$ is independent of the surface normal, it changes proportionally to the surface area when the shape of the surface is modified. For example, by ``roughening" the spherical surface to $R(\varphi)=R_0(1+\epsilon \sin m\varphi)$, the surface integral scales as $\sqrt{1+\alpha \epsilon^2 m^2}$, where $\alpha$ is a geometrical factor of order one, weakly dependent on the argument $\epsilon m$. As a result, $N_{\rm LK}$ can take arbitrary values, adjustable by the modulation amplitude $\epsilon$ and the spatial frequency $m$. At the same time, each piece of surface term in the exact normalization is proportional to the flux of ${\bf k}$, as clear from \Eq{surf}, and thus independent of the surface roughness.

Finally we show that the claim in \cite{KristensenPRA15}, that the LK normalization is equivalent to the PML normalization of~\cite{SauvanPRL13}, is incorrect. This should be clear considering that $N_{\rm LK}$ diverges, while the PML normalization is finite, as demonstrated in the supplement of~\cite{SauvanPRL13} for the RS shown in \Fig{fig:SP}. The PML normalization uses a PML to convert the radiation losses into absorptive losses within the PML, such that the remaining radiation losses at the external border of the PML can be neglected.

The equivalence of the LK and PML normalization is shown in \cite{KristensenPRA15} analytically, using the Silver-M\"uller radiation condition. This condition states that the vector field
\be
{\bf F}=\frac{\bf r}{r}\times\nabla\times {\bf E}+ik {\bf E}\,,
\ee
vanishes at large distances from the optical system, i.e. ${\bf F}\to 0$ as $r\to\infty$. Here ${\bf E}$ is the electric field of a wave emitted from the system centered at the origin, with a wave vector $k$ which is {\em real} and positive~\cite{Martin06}. However, for a RS,  $k$ is typically {\em complex}, so that the Silver-M\"uller  condition does not hold, and a divergence ${\bf F}\to \infty$ as $r\to\infty$ is found instead. To exemplify this, we take TE vector spherical harmonics, which can be used, along with their TM counterparts, for expansion of any mode of a finite system in the outside area. Their field can be written as
\be
{\bf E}=-{\bf r}\times \nabla f\,, \ \ \ {\rm where} \ \ \ f({\bf r})= h_l(kr)Y_{lm}(\theta,\varphi)\,,
\ee
so that
\bea
{\bf F}&=&\frac{\bf r}{r}\times\left [2-ikr+({\bf r}\cdot \nabla)\right]\nabla f
\nonumber\\
&=&\frac{h_l(kr)}{2ikr^2}\,\frac{P'_l(\xi)}{P_l(\xi)} \left( {\bf e}_\varphi\partial_\theta -{\bf e}_\theta\frac{\partial_\varphi}{\sin\theta}\right) Y_{lm}(\theta,\varphi),\;
\eea
in which ${\bf e}_\varphi$ and ${\bf e}_\theta$ are the unit vectors of the spherical coordinate system, and $\xi=(-2ikr)^{-1}$. We see that ${\bf F}$ diverges for $r\to\infty$ due to the exponentially growing factor in $h_l(kr)$ and the non-vanishing factor $P'_l(\xi)/P_l(\xi)\to l(l+1)$. Using ${\bf F}\to 0$ for $r\to\infty$ in Eq.\,(17) of~\cite{KristensenPRA15}, the authors obtain the LK normalization from the PML normalization. This shows actually that the two normalizations differ by a term proportional to ${\bf F}$ which is diverging for $r\to\infty$, consistent with the fact that the LK normalization is diverging while the PML normalization is not.

\acknowledgments This work
was supported by the Cardiff University EPSRC Impact Acceleration Account EP/K503988/1 and the S\^er Cymru National Research Network in Advanced Engineering and Materials.


\end{document}